\documentclass[review]{elsarticle}

\usepackage{lineno,hyperref}
\usepackage{amsmath,amsfonts}
\usepackage{booktabs}
\usepackage{float}
\usepackage{multirow}
\usepackage{subcaption}
\usepackage{makecell}
\usepackage{xcolor}

\modulolinenumbers[5]

\journal{Computer Methods and Programs in Biomedicine}

\biboptions{sort&compress}

\ifpdf
    \graphicspath{{Figs/Raster/}{Figs/PDF/}{Figs/}}
\else
    \graphicspath{{paper/Figs/Vector/}{paper/Figs/}}
\fi
\usepackage{atbegshi}
\AtBeginDocument{\AtBeginShipoutNext{\AtBeginShipoutDiscard}}

\let\oldeq\equation{}\def\equation{\par\vspace{-\parskip}\oldeq}

\begin{document}
\begin{frontmatter}

\title{Haemodynamic analysis using multiphase flow dynamics in tubular lesions}
\author[mymainaddressA]{Konstantinos G. Lyras\corref{mycorrespondingauthor}},
\address[mymainaddressA]{School of Biomedical Engineering $\&$ Imaging Sciences, King's College London, London SE1 7EU, United Kingdom}\ead{konstantinos.lyras@kcl.ac.uk}\cortext[mycorrespondingauthor]{Corresponding author}
{Corresponding author}
\author[mymainaddressA]{Jack Lee \corref{mycorrespondingauthor}}\ead{jack.lee@kcl.ac.uk}

\begin{abstract}
\textit{Background and Objective}: {\color{black} The role of red blood cell dynamics is emphasised in certain cardiovascular diseases, and thus needs to be closely studied. A multiphase model of blood flow allows the resolution of locally varying density of red blood cells within a complex vessel geometrical domain, and haemodynamic consequences of such build up.}

\textit{Methods}: A novel computational fluid dynamics solver for simulating multiphase flows is used for modelling blood flow using level set for a sharp interface representation. Single-phase simulations and reduced order models are used for pressure comparisons. {\color{black}The new solver is used for numerically studying AHA type B lesions. The impact of hematocrit and degree of stenosis on the haemodynamics of coronary arteries is investigated.} 

\textit{Results}: The comparisons with single-phase flow simulations indicate differences in pressure when considering red blood cell aggregation.
Multiphase simulations provide slightly lower pressure drop for the same stenosis severity compared to the single-phase simulations.
Secondary flow patterns and the interactions between the two phases leads to the red blood cell aggregation at the end of the diastole cycle, which significantly changes the red blood cell distribution, the shear stresses and velocity in tubular lesions.

\textit{Conclusions}: {\color{black} Neither pressure drop nor mean velocity are not strongly changed in the multiphase modelling, but particle buildup significantly changes which is only revealed by the multiphase approach.}
\end{abstract}

\begin{keyword} 
red blood cell; stenosis; coronary artery; multiphase mixture model; level set; wall shear stress
\end{keyword}

\end{frontmatter}


\section{Introduction}
Blood is a multi-component fluid primarily consisting of dispersed red blood cells (RBCs), white blood cells (WBCs) and platelets, in plasma. Normal levels of the volume fraction of RBCs (hematocrit) range from 35$\%$ to 50$\%$ \citep{reinhart2016optimum}. Consequently, the rheological behaviour of RBCs strongly affect blood properties which significantly change when the hematocrit level varies for instance in hemodilution-increased plasma/anemia-reduced RBC \citep{nader2019blood}. Particle aggregation strongly depends on hematocrit and can lead to large fluctuations of shear rate locally which effects blood viscosity \citep{mehri2018red}. This non-Newtonian shear-thinning rheological behaviour of blood can have a significant impact on blood flow \cite{yin2013multiple}. This manifests that a single fluid approach with uniform properties for numerically investigating cardiovascular diseases such as atherosclerosis might fail to model the underpinning phenomena due to particulate motion.

A multiphase computational fluid dynamics (CFD) model of blood flow can be used for the accurate resolution of locally varying distribution of RBC within a vessel and the haemodynamic consequences of such build up. The majority of continuum multiphase CFD studies for multi-component blood flow modelling are based on Eulerian-Eulerian approaches where a simple transport equation for the volume fraction of each phase is solved. These Eulerian approaches offer the opportunity of calculating the forces exerted on the fluids by solving for momentum for each phase allowing for studying inter-phase mass and momentum transfer. 
Some studies choose to couple the Eulerian-Eulerian framework with implicit methods for interface capturing, leading to a more accurate calculation of the volume fraction for each phase \cite{kim2016computational}. These are usually based on the volume of fluid (VOF) method \cite{Hirt1981}. 
VOF is one of the most commonly used implicit methods which combines simplicity and mass conservation by default. Given a sufficient grid resolution, the VOF approach models the larger scales and bulk fluid-fluid interfaces or for cases where interdispersion between two fluids can be neglected \cite{Prosperetti2009}.
The correct calculation of the interface curvature and conservation of mass for the RBC component while maintaining numerical stability in regions where volume fraction exhibits sharp changes are extremely important for accurate multiphase CFD models for numerically investigating the behaviour of RBC and plasma in cardiovascular diseases such as atherosclerosis.
{\color{black} Increasing the accuracy of interface capturing for the larger scales is a key point for increasing overall accuracy of the multiphase analysis. 
For addressing these challenges here, we deploy more advanced implicit methods rather than the traditional VOF, while guaranteeing that mass is conserved.}

In this paper we propose an improved numerical approach for the calculation of the volume fraction of each phase for modelling the RBC distribution. For this purpose, we employ a coupled VOF and level set method that we have recently developed and validated for different multiphase flow cases in \cite{LyrasLee2021a} and has been proven to give a more accurate interface description compared to other state-of-the-art VOF-based approaches.
Blood is modelled as a mixture and the interface of the two phases is captured using one of the most attractive features of the level set e.g a sharp interface representation. The formulation leads to a multi-scale approach that combines the advantages of both level set for the larger scales, and Eulerian-Eulerian two-fluid approach which can be used to model the characteristics of the mixture into the under-resolved (sub-grid) scales.  

The presented methodology is tested for the study of partially occluded arteries due to a stenotic obstruction of eccentric shape. Although these lesions account for the majority of lesions among patient population \cite{hausmann1994lumen, waller1989eccentric} they are given little attention in literature from numerical works. Contrary to symmetric stenosis models, asymmetry not only offers a more realistic representation of diseased arteries, but the generated secondary flow patterns alter the flow downstream at the end of the stenosis, resulting in a fundamentally different haemodynamics compared to the idealised axisymmetric stenoses \cite{kaazempur2005mass}.
{\color{black} In addition to eccentricity, the length of a lesion, $L_{s}$ is another important factor that also affects blood flow \cite{LyrasLee2021c}}. Although some studies for discrete lesions with $L_{s} \le 10mm$ --type A in the American College of Cardiology/American Heart Association (ACC/AHA) classification of coronary lesions, have been studied \cite{gholipour2018, zhao2021}, there are no detailed studies for tubular coronary lesions (type B, with $ 10mm \le L_{s} \le 20mm$) which involve a moderate risk, higher than in discrete lesions \cite{kushner2009}.
To the best knowledge of the authors, the effects of hematocrit and stenosis on the RBCs transport and its effect on the blood flow in tubular coronary lesions (with $ 10mm \le L_{s} \le 20mm$, $L_{s}$= lesion length) have not been previously reported elaborately. 
In this study, the blood flow patterns and particulate aggregation {\color{black}in coronary arteries with a tubular lesion} are numerically studied using the developed multiphase solver.

{\color{black}Owing to the complexity of the multiphase flow phenomena and their dependence on the lesion geometry, apart from the eccentricity and length of a lesion, the hematocrit is also studied for investigating its role on the local haemodynamic parameters such as the velocity and the WSS on the endothelial cell membrane in the coronary artery.}
Although the effects of the hematocrit in the viscosity have been previously studied \cite{benis1970effect, ccinar1999effect}, many open questions still remain regarding the local haemodynamic factors, such as the wall shear stress, red blood cell buildup and velocity \cite{jung2006hemodynamic, kim2008multiphase, baskurt2011red, lucker2015dynamic, ling2021numerical}. These important parameters which vary alongside the curved vessel and during the cardiac cycle \citep{Thim2012, Han2016} are the primary focus of this research here.
First, the impact of hematocrit and stenosis severity on the calculated pressure drop $\Delta p$ is studied.  
Calculating $\Delta p$ along a lesion is extremely important since it is typically used for characterising the significance of the stenosis through the calculation of Fractional Flow Reserve (FFR). FFR is a haemodynamic index, based on invasive measurement of the pressure ratio distal and proximal to a given lesion which is employed for guiding revascularisation.  
Since in the particular case of tubular lesions, there are no numerical studies that evaluate the effects of the RBC levels on $\Delta p$, the velocity and red blood cell distribution, we include some tests together with comparisons with single phase and reduced order models.

Comparisons with single-phase flow demonstrate the differences in pressure when considering the mixture.
Single-phase simulations over-predict the pressure drop for the same stenosis severity compared to the multiphase simulations. The model here generally predicted higher RBC buildup in the area of maximum curvature and the end of the stenosis compared to the upstream location.
WSS region was found to vary with respect to the local RBC buildup, in the area of maximum curvature of the vessel. Secondary flow patterns and vessel geometry significantly influence the RBC buildup, primarily because of the prolonged particulate residence times. 
These observations could have significant implications towards better understanding of the haemodynamic parameters that might have an impact on the development of cardiovascular diseases such as atherosclerosis in tubular coronary lesions.
 
\section{Materials and Methods}
\label{sec:Euler-Euler framework}
The open-source code OpenFOAM \cite{weller1998} is used for the simulations presented. The numerical implementation here has used as its basis the solver multiphaseEulerFoam solver \cite{wardle2013hybrid} and enables simultaneous modelling of complex multiphase flows with different flow regimes (i.e. both dispersed and segregated flows). 
The solver uses a coupled level set (LS) and VOF interface capturing method to model the large fluid-fluid interfaces, whereas the Euler multi-fluid approach is used for modelling the dispersed flow. The two approaches are coupled by a simplified switching term, which is defined for each phase pair \cite{wardle2013hybrid}. 
The solver is a combination of two multiphase solvers i.e. a coupled VOF-LS and an Eulerian-Eulerian solver with additional sub-models to model the forces involved.  
In addition, the solver includes models for phase change (e.g. cavitation) which is deactivated here.  
Although the methodology can be used to simulate multiple fluids, here we choose to present it with two fluids. This is an Eulerian-Eulerian two-fluid model which regards RBCs as suspended particles dispersed in plasma, which has been shown that it accurately describes the blood flow \cite{yin2013multiple}. 
Following other researchers, the effect of other components such as platelets was considered here to be insignificant, since they exhibit volume fraction values which are less than 1$\%$     
\cite{jung2006hemodynamic}.
The multi-fluid model equations for incompressible, isothermal flow are given by a set of mass and momentum equations for each phase k,

\begin{equation}
\frac{\partial (\rho_k \alpha_k u_k)}{\partial t}+ (\rho_k \alpha_k u_k) \cdot \nabla(u_k)=-\alpha_k \nabla p + \nabla \cdot (\mu_k \alpha_k \nabla u_k) + \rho_k \alpha_k \mathbf{g} + \mathbf{F}_{D,k}+ \mathbf{F}_{s,k} + \mathbf{F}_{vm}
\label{momentumEqn}
\end{equation}
where the density, phase fraction, and velocity for phase $k$ are given by $\rho_k$, $\alpha$, $u_k$. The $\rho_k \alpha_k \mathbf{g}$ is the gravitational force and the other two interfacial forces in Eq.\ref{momentumEqn} are the drag force $\mathbf{F}_{D,k}$ and the surface tension force $\mathbf{F}_{s,k}$. For selected phase pairs where the VOF method is used to resolve the interface, an additional compression term is added to their phase transport equation. The method here includes the level set method for capturing the interface of each pair of fluids.
The drag term $\mathbf{F}_{D,k}$ is given by 
\begin{equation}
\mathbf{F}_{D,k} = \frac{3}{4}\rho_c \alpha_c \alpha_d C_D \frac{|\mathbf{u}_d - \mathbf{u}_c|(\mathbf{u}_d - \mathbf{u}_c)}{d_d}= \alpha_c \alpha_d K (u_d - u_c)
\end{equation}
where the subscripts $c$ and $d$ denote the continuous (plasma) and dispersed phase (RBC) values respectively and $K$ is 
\begin{equation}
K = \frac{3}{4} \rho_c C_D\frac{|\mathbf{u}_d - \mathbf{u}_c|}{d_d}
\end{equation} 
Among the various models for calculating the drag coefficient $C_D$ available in OpenFOAM, the commonly used model of Schiller and Naumann \cite{schiller1933drag} is employed here. In this model, the drag coefficient is a function of the Reynolds number $Re$ according to
\begin{equation}
C_D = 
\begin{cases}
\frac{24(1+0.15Re^{0.683})}{Re} & \text{, } Re \leq 1000 \\
0.44 & \text{, } Re > 1000  
\end{cases}
\end{equation}
where the Reynolds number is calculated based on the relative velocity of the two phases, the dispersed phase diameter and the kinematic viscosity $\nu$ according to
\begin{equation}
Re = \frac{|\mathbf{u}_d - \mathbf{u}_c |d_d}{\nu_c}
\end{equation}
Calculation of the drag coefficient can be done by the independent calculation with each phase as the dispersed phase and the overall drag coefficient applied to the momentum equations taken as the volume fraction weighted average of the two values. 
This blended scheme is a useful approximation for flows with regions in which either phase becomes the primary phase.
The virtual mass effect accounts here for the acceleration of the particulate phase (RBC) relative to the plasma. The virtual mass forces $F_{vm,c},F_{vm,d}$ are calculated for both phases as
\begin{equation}
\mathbf{F}_{vm,c} = -\mathbf{F}_{vm,d} = 0.5 \alpha_d \rho_c \left( \frac{\text{d}u_c}{\text{d}t} -\frac{\text{d}u_d}{\text{d}t}\right)
\end{equation}
The lift force acts on the RBCs due to the developed velocity gradients in the plasma which has its own velocity field. The lift forces are calculated as 
\begin{equation}
\mathbf{F}_{lift,c} = -\mathbf{F}_{lift,d} = -0.5 \alpha_d \rho_c \left(\mathbf{u}_c -\mathbf{u}_d\right) \times (\nabla \times\mathbf{u}_c ) 
\end{equation}
A constant droplet diameter size, defined independently for each phase is used in this framework with 
the RBCs considered as spherical particles with a diameter of 8 $\mu$m
\cite{jung2006hemodynamic}.

\label{sec:Level set method}
\subsection{Marker function for level set}
The level set function is used here to describe the interface dynamics of the fluids. For simplicity, the case of two fluids is presented. The level set function $\psi$ is defined to be a signed distance function calculated from the distance at the interface, $\Gamma$ that separates the two fluids 
\begin{equation}
|\psi| = min(\vec{x}-\vec{x_{\Gamma}})
\label{psiDefinition}
\end{equation} 
A given point $\vec{x}$ can have either a positive or a negative distance which defines whether it belongs in one fluid or the other. The interface that separates the two fluids is then the set of points that have a zero-distance (zero-level). 
A Heaviside function $H(\psi,t)$ is used to calculate the smoothed volume fraction that depends on the values of $\psi >0$. Here, the following expression is chosen \cite{Olsson2005},
\begin{equation}
H(\psi) 
= \begin{cases}
1 & \text{\quad if in fluid 1}  \\
\frac{1}{2}\left[ 1+\frac{\psi}{\epsilon}+\frac{1}{\pi}sin(\frac{\pi \psi}{\epsilon})\right] 
 & \text{\quad at the interface} \\
0 & \text{\quad if in fluid 2}  
\end{cases}
\label{HEqn}
\end{equation}  
In order to conserve mass, the equation for $H$ needs to be satisfied 
\begin{equation}
\frac{\partial H(\psi,t)}{\partial t}+ \mathbf{u} \cdot \nabla{ H(\psi,t)} =0
\label{rhoHEqn}
\end{equation}
Then $H(\psi,t)$ resembles a smoothed volume fraction $\alpha_c$ of one fluid and 1-$H(\psi,t)$ for the other fluid. The total mass with respect to time is then $\int_{V}^{}\rho_i H(\psi)dV$.
The level set function according to Eq.\ref{psiDefinition} is a distance function that is defined wherever an interface exists. The distance function can be advected using 
\begin{equation}
\frac{\partial \psi}{\partial t}+ \mathbf{u} \cdot \nabla{\psi} =0
\label{psiEqn}
\end{equation}
Eq.\ref{psiEqn} is solved first and the distance function is advected. The level set is then re-initialised for allowing $\psi$ to remain a signed distance function. For this, the Eikonal equation $|\nabla \psi| = 1$ is solved. 
\begin{equation}
\frac{\partial \psi_{d}}{\partial 
\tau}=\textrm{sgn}(\psi)(1-|\nabla \psi_{d}|)
\label{reinitGeneric}
\end{equation}
The solutions of Eq.~\eqref{reinitGeneric} have the same zero-level as $\psi$ and are distance functions. 
Denoting with $\tilde{\psi}_d$ e.g. $\tilde{\psi}_d=\psi^{n+1}_d$ the new value of $\psi$ after solving Eq.~\eqref{reinitGeneric} an additional step for satisfying mass conservation is performed. \cite{LyrasLee2021a}
The correction denoted here with $\hat{\psi}$ is added to $\tilde{\psi}_d$ so that the following is satisfied for an appropriate $\psi$ \citep{Kees2011}
The following reaction-diffusion equation is solved 
\begin{equation}
H(\psi_d + \hat{\psi}) = \alpha_c + \lambda \Delta \psi_d 
\label{massConservationCorr}
\end{equation}
where $\lambda$ is a parameter that is a function of the cell size and is generally taken here $\approx 0.5(\Omega)^{1/3}$, where $\Omega$ is the volume of the cell. 
Eq.~\eqref{massConservationCorr} is solved with an implicit solver with the laplacian term using an unbounded, second-order conservative scheme using linear interpolation. The resulting $\hat{\psi}$ value leads to the final level set value  
\begin{equation}
\psi^{n+1} = \psi^{n+1}_d + \hat{\psi}
\end{equation}
This new value satisfies the mass conservation since $H(\psi^{n+1}) = \alpha_c$ and preserves on of the most attractive features of LS methods, a sharp interface profile wherein an accurate curvature value is obtained and is directly used in the momentum equation Eq.\eqref{momentumEqn}. 

The method avoids the assumption made by similar approaches for the interface position that is not located at the iso-surface contour $\alpha$-0.5 as in \citep{Kunkelmann2010, Albadawi2013, Dianat2017, Lyras2020}. Compared to other previous works, the level set function is not reconstructed only from the advected volume fraction \citep{Albadawi2013, Lyras2020} but neither is based on specifically designed schemes for re-initialisation \cite{Lyras2020, Hartmann2010} that allows for simpler and faster calculations.

\subsection{Switching between segregated and dispersed flow regimes}
First the transport equation for obtaining an initial approximation of the liquid volume fraction is solved. The equation includes an interface compression term $\mathbf{u}_c$ and is given as
\begin{equation}
\frac{\partial \alpha_c}{\partial t}+ u_k \cdot \nabla(\alpha_c) + \nabla \cdot (u_c \alpha_c (1-\alpha_c))=0
\end{equation}
where the velocity $u_c$ is applied normally to the interface to compress the volume fraction field and maintain a sharp interface. The $\alpha_c (1-\alpha_c)$ term ensures the term is only active
in the interface region. The value for the artificial interface compression velocity is given by

\begin{equation}
u_c = min (C_{\alpha}|u|,max(|u|))\frac{\nabla \alpha_c}{|\nabla \alpha_c|}
\end{equation}
The $\frac{\nabla \alpha_c}{|\nabla \alpha_c|}$ term is the interface unit normal vector
for the direction of the applied compression velocity.   
The coefficient $C_{\alpha}$ is used for controlling the interfacial compression, and when considered to be $\in [0,1]$ it is simply a binary coefficient which switches interface sharpening on/off when it is equal to 1/0 respectively. 
When $C_{\alpha}=0$ for a given pair of phase, no interface compression is imposed resulting in phase dispersion. When $C_{\alpha}=1$, sharp interface capturing is applied and the flow is treated with VOF.

The switch between approaches is done dynamically imposing the interface sharpening in those areas of the domain that the $C_{\alpha}$ disctates. This allows for a unified method which allows simulation of complex flows with any combination of regimes ranging from fully dispersed to fully segregated. The switch is done based on evaluating the magnitude of the gradient of the volume fraction ($\gamma$), assuming that when the gradient of the volume fraction becomes smaller than a defined cut-off value ($\gamma_{cut-off}$) phase dispersion occurs and $C_{\alpha}=0$. The parameter $\gamma$ is then defined by\cite{wardle2013hybrid}
\begin{equation}
\gamma = \frac{\nabla \alpha_c}{max(\nabla \alpha_c)}
\end{equation}
Thus, when $\gamma \geq \gamma_{cut-off}$ the value of $C_{\alpha}$ is set to 1. The value of $\gamma_{cut-off}$ is set to 0.4. 

\section{Results}
\label{sec:results}
\begin{figure}[h]
 \vspace{6pt}
\centering
\begin{subfigure}{.58\textwidth}
  \centering
  \includegraphics[width=1.\linewidth]{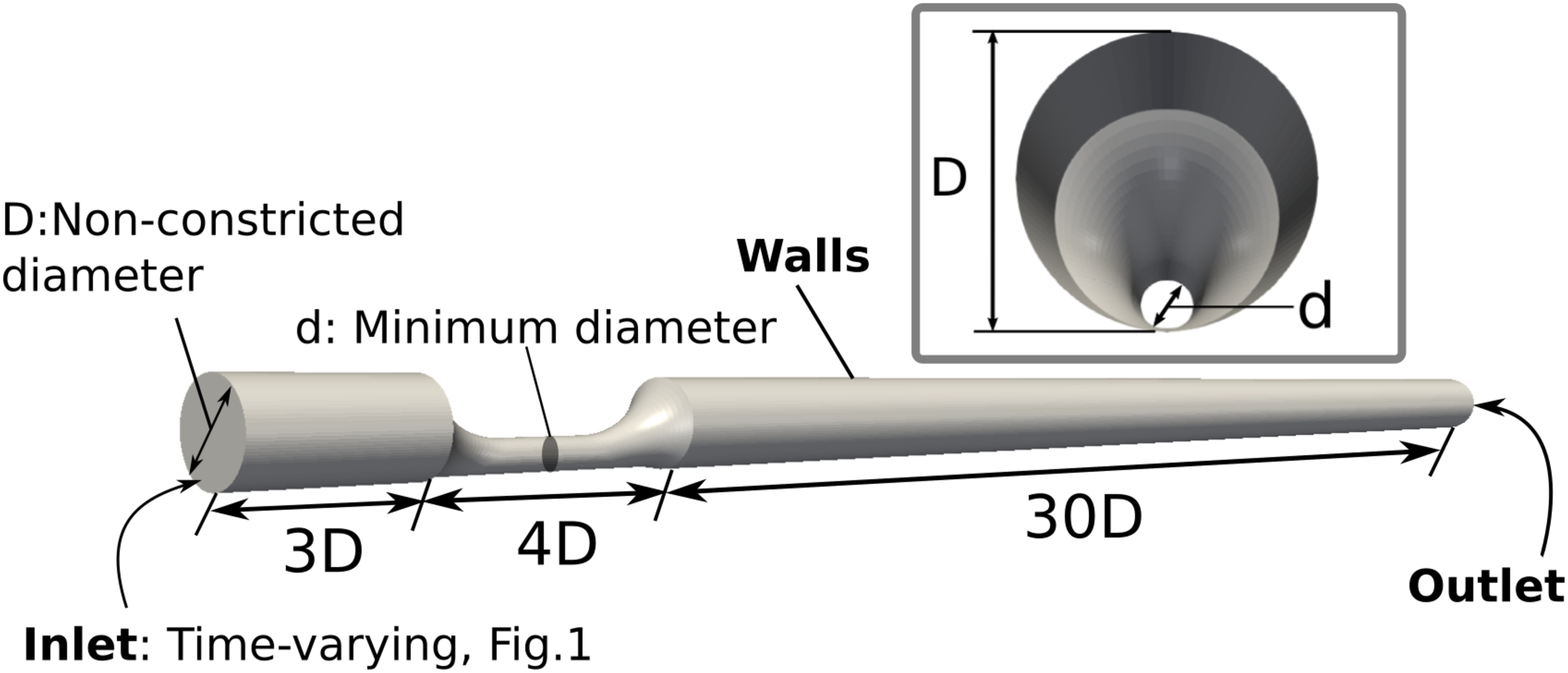}
  \label{fig:sub1}
\end{subfigure}%
\begin{subfigure}{.58\textwidth}
  \centering
  \includegraphics[width=1.\linewidth]{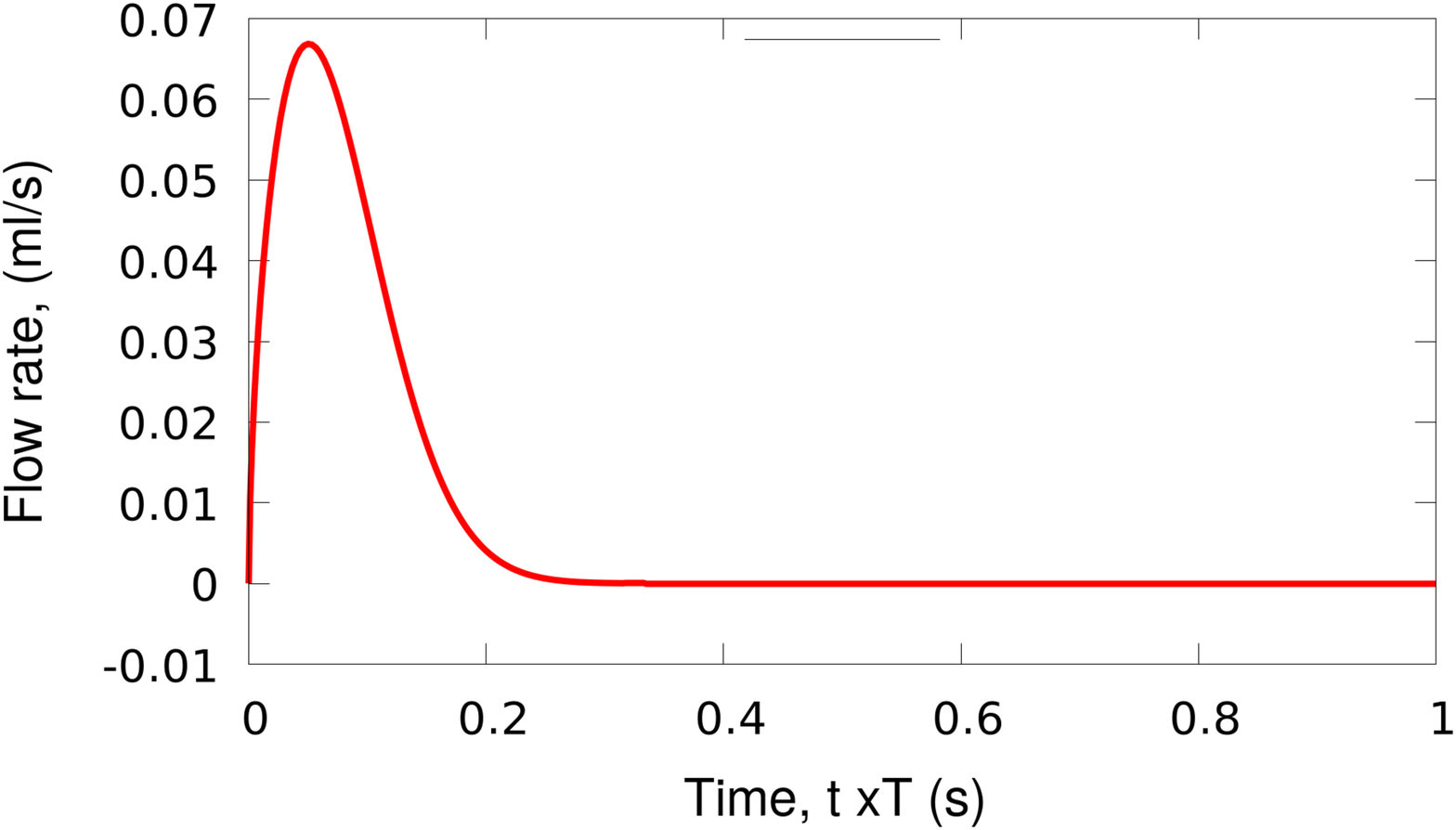}
  \label{fig:sub2}
\end{subfigure}
 \caption{Domain and inlet used here for the numerical tests.}
    \label{fig:geoms_bc2}
\end{figure}

The numerical tests here included an eccentric stenosis with a stenosis severity $S_{D}$ that varied from $50\%$ to $70\%$. Two cases of different hematocrit levels are used, one with RBC=35$\%$ and another one with RBC=45$\%$. The length of the stenosis $L_{s}=14 mm$ and the non-constricted diameter of the artery was $D=4mm$, shown in Fig.\ref{fig:geoms_bc2}. 
For the inlet velocity, a pulsating unsteady blood flow with a period of $T=1s$ was imposed and is shown in Fig.\ref{fig:geoms_bc2}. The flow upstream the stenosis was assumed to be fully developed downstream with a Neumann boundary condition considered at the outlet. A zero slip velocity was employed for the RBCs and plasma at the lesion walls. The inlet profiles for velocity and RBC were considered to be uniform across the inlet cross section similar to other similar works\cite{jung2006hemodynamic}. 

The multiphase CFD results for pressure drop were compared with the results obtained from two reduced-order models (ROMs). 
The model of Young and Tsai\cite{young1973steady,young1973unsteady} is a trans-stenotic pressure drop model developed for general vascular stenosis. The model uses some empirical parameters obtained from in vitro experiments. 
The model of Lyras and Lee \cite{LyrasLee2021b} was developed based on analysis of the phenomena that contribute the most to pressure reduction in coronary arteries e.g. the viscous and the inertial effects, turbulence and unsteady flow. 
These models treat the stenosis as one-dimensional fluid flow through an area-varying tube. Although these models provide a simplified expression for pressure drop, they have been shown to be to predict $\Delta p$  with reasonably small difference compared to CFD and experiments.\cite{stergiopulos1992,LyrasLee2021b} Together with the results from the reduced models, the results here are compared to the case of single-phase CFD simulations with laminar and Reynolds-Averaged Navier-Stokes(RANS) simulations are considered for the tests here. The $k-\omega-$SST model \cite{menter1993} is used here and has been shown to be an appropriate choise for the flows considered here \cite{khair2015}. 
   
\begin{figure}[h]
    \vspace{6pt}
    \centering
    \includegraphics[scale=0.18]{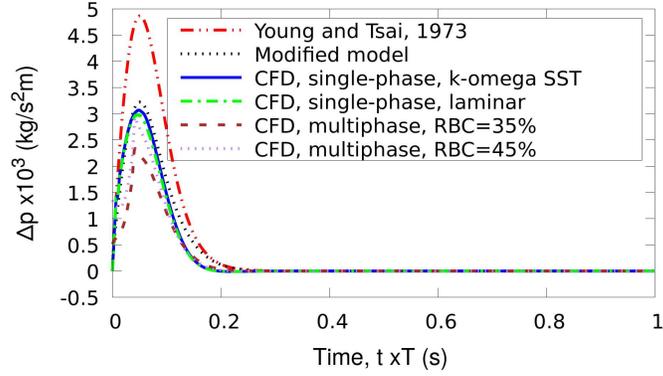}
       \centering
    \caption{Pressure drop inside the vessel for different RBC and for 50$\%$ stenosis severity. Results are compared with the reduced-order models and the single phase flow approach.}
    \label{fig:pressure_50pc}
\end{figure}

\begin{figure}[h]
    \vspace{6pt}
    \centering
    \includegraphics[scale=0.18]{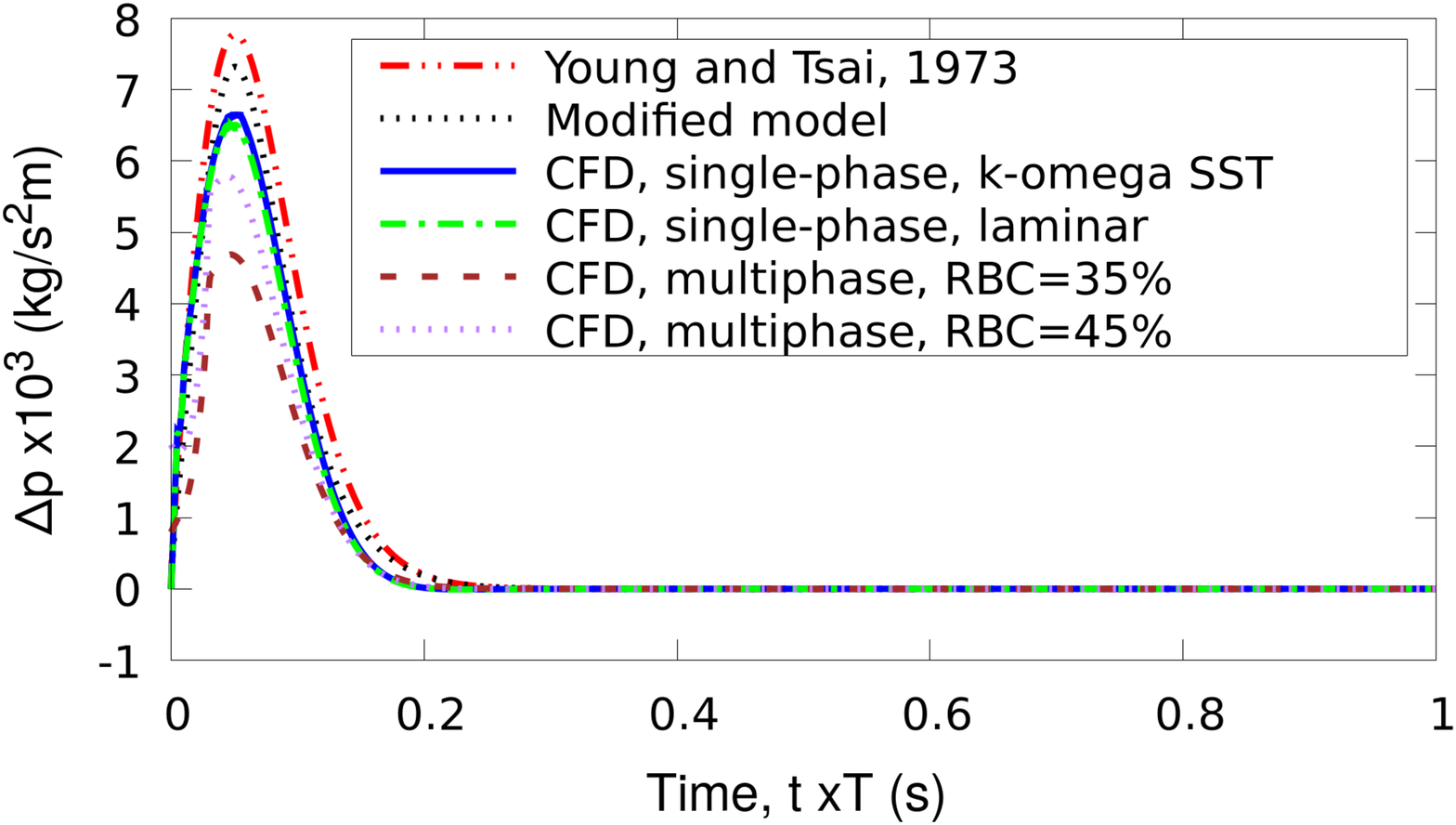}
       \centering
    \caption{Pressure drop inside the vessel for different RBC and for 70$\%$ stenosis severity. Results are compared with the reduced-order models and the single phase flow approach.}
    \label{fig:pressure_70pc}
\end{figure}

\begin{figure}[H]
 \vspace{6pt}
\centering
\begin{subfigure}{.58\textwidth}
  \centering
  \includegraphics[width=1.\linewidth]{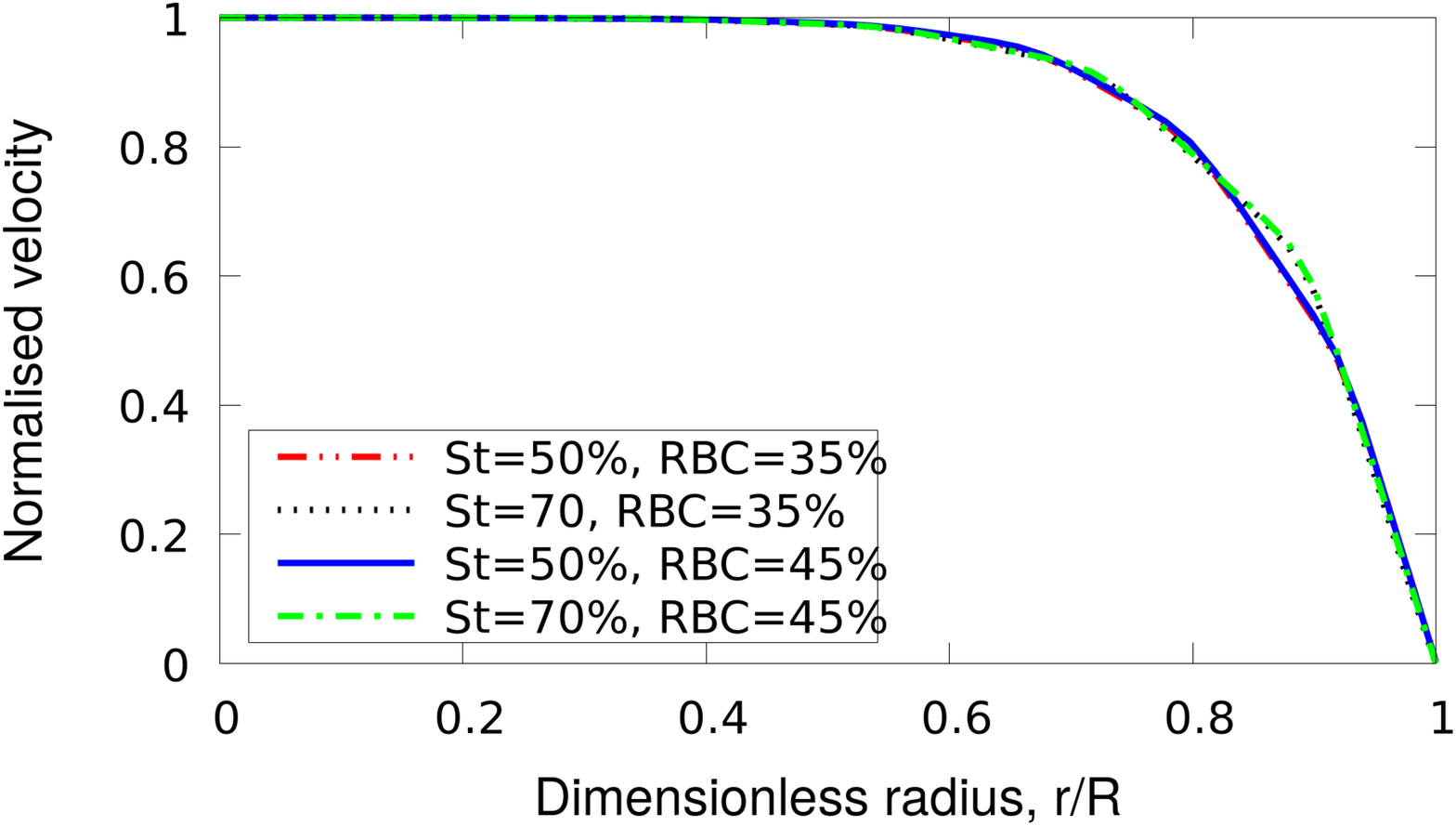}
  \label{fig:sub1}
\end{subfigure}%
\begin{subfigure}{.58\textwidth}
  \centering
  \includegraphics[width=1.\linewidth]{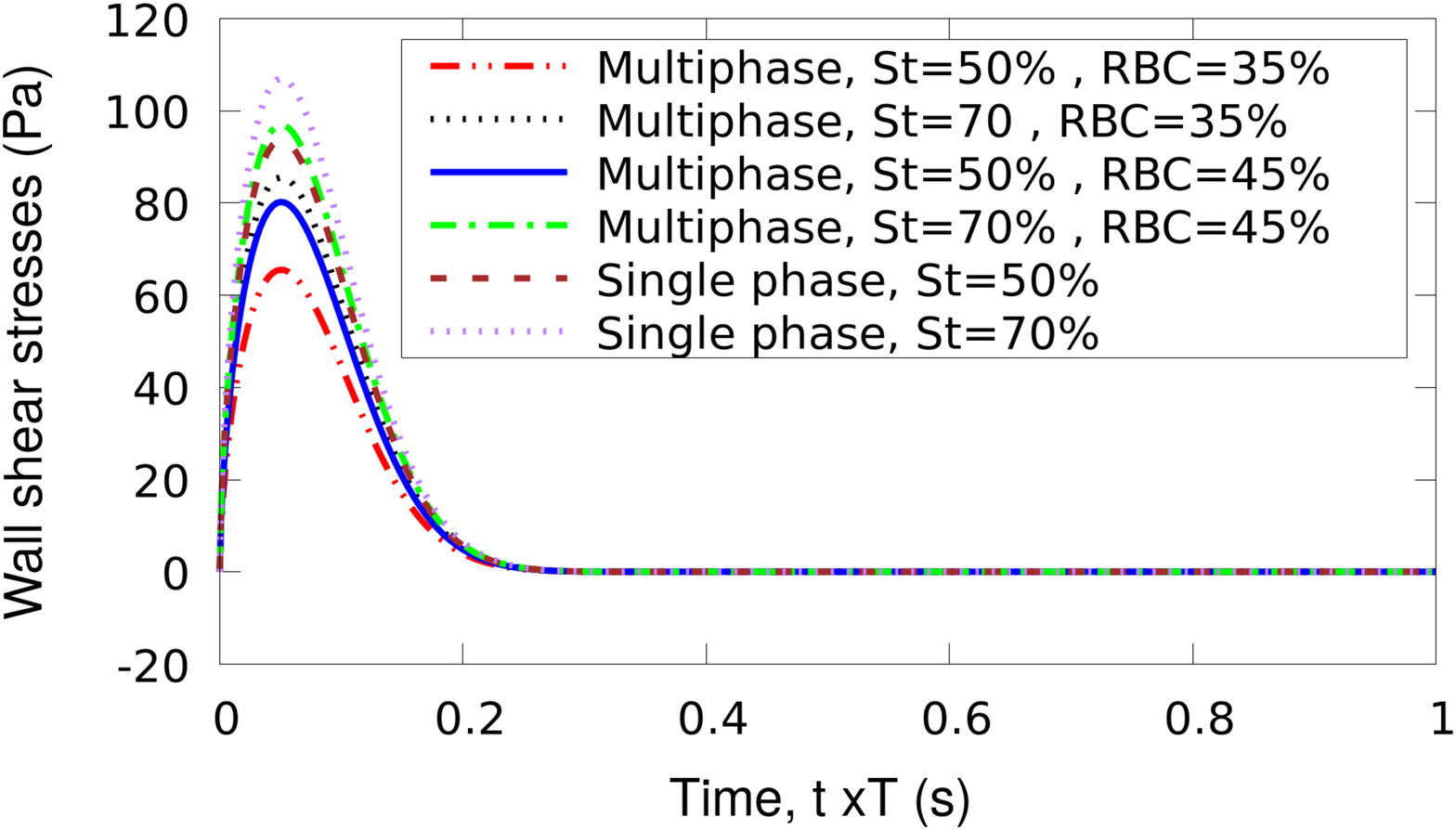}
  \label{fig:sub2}
\end{subfigure}
\caption{Velocity distribution at the radial direction and the x=L2 position (on the left). The shear stresses are shown with respect to time at the point of maximum wall shear stress at x=L2 (on the right).}
\label{fig:ux_wss}
\end{figure}

\begin{figure}[H]
    \vspace{6pt}
    \centering
    \includegraphics[scale=0.21]{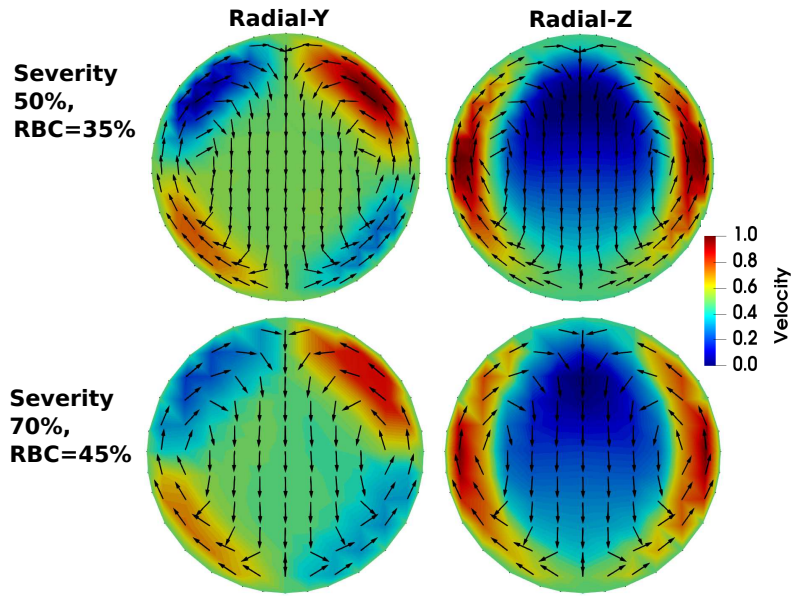}
       \centering
    \caption{Normalised velocity for the two different cases for different RBC and for 50-70$\%$ stenosis severity. Results are compared at the position L2 for the two radial directions Y-Z perpendicular to the main X flow direction.}
    \label{fig:velocity_captions}
\end{figure}

\begin{figure}[H]
    \vspace{6pt}
    \centering
    \includegraphics[scale=0.2]{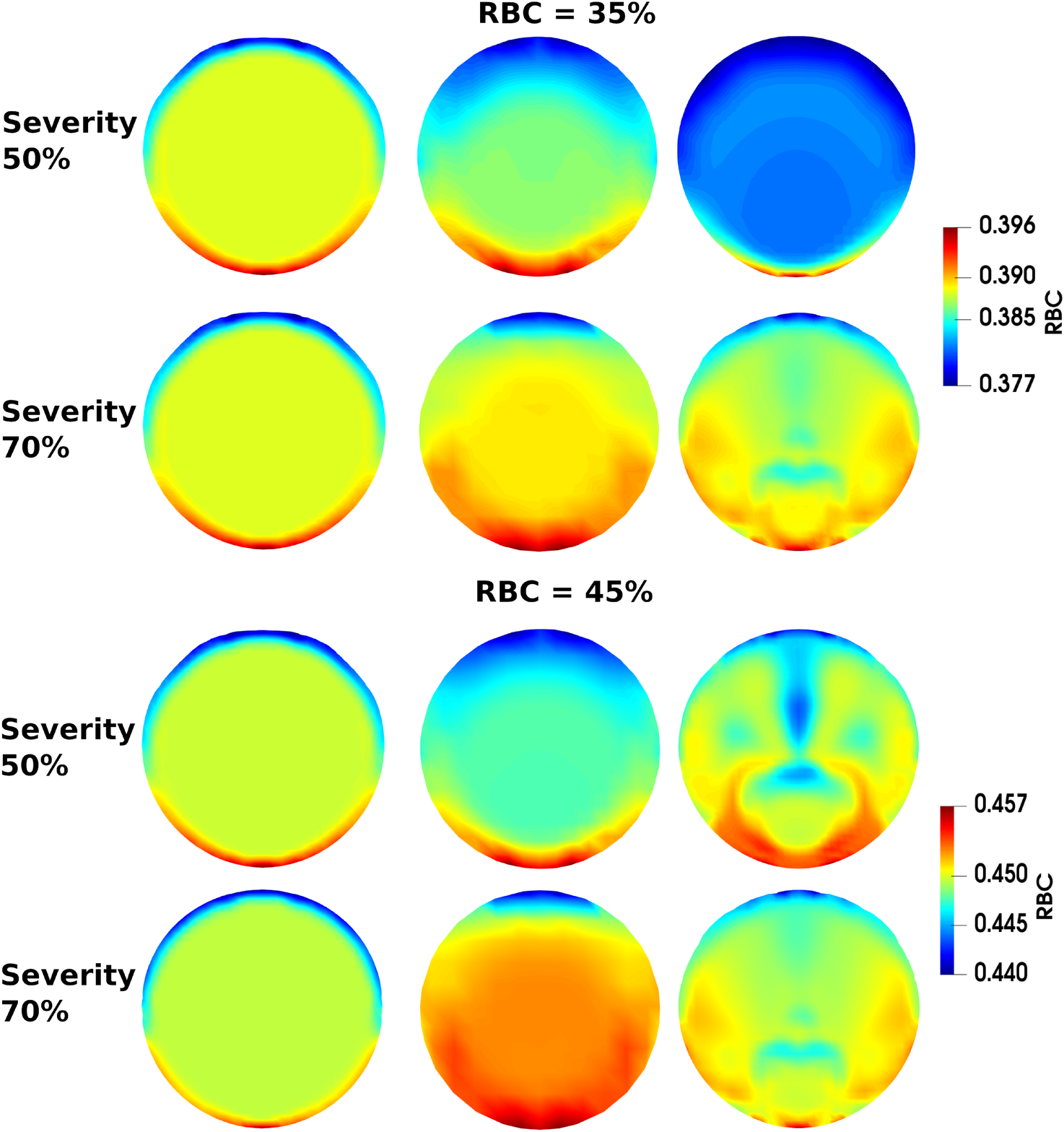}
       \centering
    \caption{Red blood cell levels for the different cases for different RBC and for 50-70$\%$ stenosis severity. Results are compared for the three different positions L1, L2, L3.}
    \label{fig:rbc_captions}
\end{figure}

\begin{figure}[H]
    \vspace{6pt}
    \centering
    \includegraphics[scale=0.16]{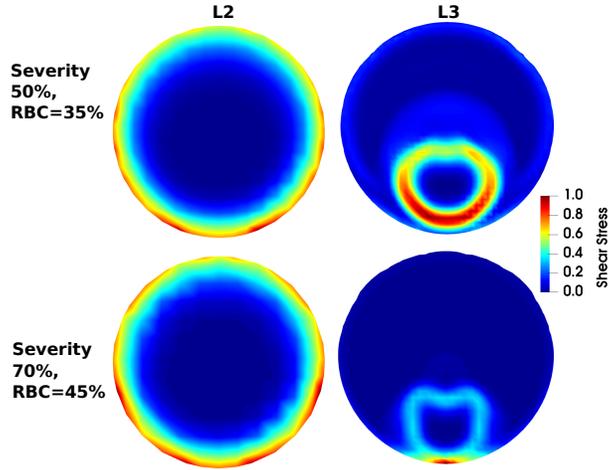}
       \centering
    \caption{Shear stresses for the different cases for different RBC and for the two cases of 50, RBC=35$\%$ and 70, RBC=45$\%$. Results are compared for the two different positions L2, L3.}
    \label{fig:wss_captions}
\end{figure}

The results for $\Delta p$ for $S_D = 50\%$ are shown first in Fig.\ref{fig:pressure_50pc} and for $S_D = 70\%$ in Fig.\ref{fig:pressure_70pc} for both RBC cases and ROMs.
The results for velocity, shear stresses and RBC are shown here for three different locations, at the beginning of the stenosis (L1), at the position at the maximum curvature at x=24mm (L2), and the end of the stenosis (L3). 
Fig.\ref{fig:ux_wss} shows the velocity distribution at the radial direction and the x=L2 position and the shear stresses with respect to time.
The normalised velocity for the two different cases for different RBC and for 50-70$\%$ stenosis severity are shown in Fig.\ref{fig:velocity_captions}. The results are compared at the position L2 for the two radial directions Y-Z perpendicular to the main X flow direction.
In Fig.\ref{fig:rbc_captions} the distribution of the red blood cells for the different cases for different RBC and for 50-70$\%$ stenosis severity are presented. Results are compared for the three different positions L1, L2, L3.
In the region of maximum curvature (L2), the volume fraction of RBC generally increased with the WSS.
Fig.\ref{fig:wss_captions} shows the shear stresses for the different cases for different RBC and for the two cases of 50, RBC=35$\%$ and 70, RBC=45$\%$. Results are compared for the two different positions L2, L3.

\section{Discussion} 
\label{sec:discussion}
Although the drag and pressure forces are the ones that primarily influence the motion of the RBCs \cite{kim2016computational}, it is not clear in the literature which if the other forces have a significant impact on the haemodynamics \cite{jung2006hemodynamic, wu2015study}. For this reason the virtual mass and lift forces are also included in the simulations together with the drag forces for describing the RBC-plasma interactions. 
Considering the effects of gravity and the main forces here is paramount for the successful simulation of the problem, the absence of which would otherwise contribute to large deviations in the RBC distribution.

The hematocrit is a key variable that determines the red blood cells aggregation which effects blood viscosity. Increasing red blood cells aggregation and consequently the viscosity, results in higher pressure drop values as shown in Fig.\ref{fig:pressure_50pc}. 

The pressure obtained with higher RBC was closer to the sigle-phase results. Since both CFD approaches use the same numerical algorithm for solving Navier-Stokes (presure-based, PIMPLE algorithm and same numerical schemes for the discritisation), this difference is attributed to the presence of RBC and the different forces included and the calculated shear stresses \cite{xu2013large}. 
The results here are in agreement with the experiments of Freidoonimehr et al.,\cite{freidoonimehr2018effect} who studied lesions with much shorter stenoses ($L_{s}/D \leq 1 $) than the ones considered here. 

The effect of hematocrit is more evident for the lesions with a higher stenosis severity as shown in Fig.\ref{fig:pressure_70pc}. For all cases, the pressure drop increases for higher $S_D$ which is expected and in accordance to previous studies and the ROM \cite{young1973unsteady, LyrasLee2021b}.
In all simulations the flow pattern changes due to the varying cross sectional area. Although the differences between the lesions with the same RBC are not significant upstream the stenosis, the flow regime is influenced by $S_{D}$ downstream (positions L2 and L3). 
An increase in the fluid velocity passing the stenosis results in the observed pressure drop which is intensified more in the tests of $S_{D}=70\%$. 
Due to the abrupt increase in the velocity, the formation of turbulent structures and flow seperation grow after the constriction leading to a more significant drop in pressure across the constrictions as observed in previous studies \cite{gould1978}.

The radial velocity profiles in Fig.\ref{fig:ux_wss} were parabolic for all cases and the effect of the hematocrit was shown to have a lesser effect on the radial velocity compared to the stenosis severity.  
The magnitude of the axial velocity were higher for both lesions with $S_D=70\%$ with a peak velocity magnitude of 1.6 m/s for RBC=45$\%$ and 0.95 m/s for $S_D=70\%$ for RBC=45$\%$ during the systole cycle. 
Our results of the velocity profile are qualitatively in agreement with the ones in Jung et al.\cite{jung2006hemodynamic} who also observed similar flat profiles at the radial direction of the vessel. 

At the L2 position the velocity profile has a parabolic-shape profile. This shape differs from the profiles reported in the work of \cite{buradi2019effect} who observed a plug-shaped flow for axisymmetric lesions of various stenosis severities including the ones studied here. 
The reason for this difference in the profile with the results here, is that in the case of the axisymmetric idealised lesions, the RBCs move towards the centreline of the vessel whereas in this study, due to eccentricity, the red blood cells exhibit a core with high RBC levels at the centreline, but also at regions near the wall at the lower part of the vessel.

The shear stresses observed with the single-phase and the two-phase approaches have the same trend with the inlet velocity. For the same stenosis, the single phase approach predicted higher values for shear stress (Fig.\ref{fig:ux_wss}-right).  
The WSS when the cross-sectional areas is smaller becomes higher, having high shear rates near the vessel wall.
The non-Newtonian shear thinning is responsible for this difference, and is related to a lower shear rate in the case the flow is considered to be multiphase. 
The approach described here treats the RBCs (which mostly contribute to the haemodynamic forces) as a shear-thinning fluid with a viscosity which is a function of the volume fraction \cite{brooks1970interactions}. 

In this study it is assumed that the red blood cells are deformable and non-spherical. Following previous  studies the dispersed phase consists of rigid spherical particles without considering the effects of the normal stress effects \cite{massoudi2012modeling}. The stress tensor for the RBCs is,
\begin{equation}
\tau_{ij,d} = k_1I + k_2S{ij,d}
\end{equation}
\begin{equation}
k_1 = -p \alpha_{d}
\end{equation}
\begin{equation}
k_2(\alpha_{d},\dot{\gamma})= \left[\mu_{\infty} + (\mu_{0}-\mu_{\infty})\frac{1+ln(1+k\dot{\gamma})}{1+k\dot{\gamma}}(\alpha_{d}+\alpha^{2}_{d})\right]
\end{equation}
\begin{equation}
\dot{\gamma} = [2tr(S^{2}_{ij,d})]^{1/2}
\end{equation}
where $\tau_{ij,d}$ is the partial stress tensor for the RBC with a corresponding shear rate $S_{ij,d}$. The constant $k_2$ accounts for the shear-thinning effect of RBCs, calculated by \cite{yeleswarapu1996evaluation}, $\dot{\gamma}$ is the RBC shear rate, $\mu_{\infty}$ is an asymptotic viscosity for the case of an infinite shear rate, $\mu_{0}$ is the viscosity for the case of zero shear rate and $k$ is a parameter that describes the shear thinning aspects of the RBCs.

The shear stresses at the L2 point in Fig.\ref{fig:ux_wss} were also higher for higher RBC concentration as shown in Fig.\ref{fig:rbc_captions} for the same stenosis, and higher than those of the single-phase model.    
Depending on the region considered inside the vessel, WSS varied with the L3 giving lower WSS in the downflow region in Fig.\ref{fig:wss_captions} compared to L2. In this downflow position, the high shear stresses regions formed in the areas of higher axial velocities and low RBC (shown in Fig.\ref{fig:rbc_captions}). The low viscosity in those areas might be also a parameter that contributes to the formation of these low-RBC regions as observed in \cite{jung2006hemodynamic}.  
For higher initial RBC levels from 35$\%$ to 45$\%$, the relative RBC volume fraction generally increased in the downflow location L3 but WSS decreased near the wall, while preserving higher values in the location L2 for both cases for different stenosis severity.

The curvature-induced secondary flow profiles exhibited vortices in the flow as shown in Fig.\ref{fig:velocity_captions}, with a recirculation pattern that varied during the cardiac cycle.
As shown in Fig.\ref{fig:ux_wss} high velocity gradients were observed at the near wall regions with a zero gradient closer to the centre.
Large secondary recirculating flows were observed in all cases here with similar patterns for the different stenosis severity and initial RBC level (Fig.\ref{fig:velocity_captions}). These recirculation zones might contribute to large pressure fluctuations locally, and lead the migration of the RBCs towards locations other than the centreline \cite{zhao2011shear}.

Fig.\ref{fig:rbc_captions} presents the distribution of the volume fraction of the red blood cells on different radial cut planes for the various severities considered when the blood flow was at maximum, i.e., at the beginning of diastole (t/T=0.05s). 

Inside the curved part of the vessel, the RBCs mitigate towards the centreline as a result of the shear-induced particulate deposition. This is more pronounced for a higher degree of stenosis and inside the maximum curvature position, since further downstream the end of the stenosis, the RBC flow pattern becomes more complicated under the influence of the secondary flow and recirculation flow motions.

At the upstream position of the stenosis (L1) a lesser RBC migration is observed with similar relative profiles for all the vessel models studied here.
On the other hand, for all the vessels, low RBC volume fraction values were generally obtained at the downstream location (L3) compared to L2, where lower WSS led to a RBC migration.

Areas with lower WSS such as areas with flow recirculation might potentially form atherosclerotic plaque \cite{ku1985pulsatile}. The RBC patterns at the downstream position were less uniform than the ones in the other locations, due the impact of turbulence, the impact of which, increases in more stenosed vessels, and was also reported in \cite{mahalingam2016numerical}.

Local fluctuations in the red blood cell distribution at the maximum curvature and downstream locations together with the changes in viscocity might be a factor that could potentially lead to a limited intake of oxygen at the arterial endothelial surface causing hypoxia \cite{ogawa1990effect}. Furthermore, nitric oxide (NO) is captured by red blood cells and scavenged inside through hemoglobin. Hence, the changes in the RBC distribution due to the particulated deposition might lead to changes in the NO transport by the red blood cell hemoglobin which has been reported to cause atherogenesis and atherosclerosis \cite{premont2020role, jeney2014red}.

The WSS is a significant haemodynamic parameter that is usually considered to study the platelet and vascular behaviour in blood flow through vessels. Areas with high WSS might have increased risk for diseases such as thrombosis. Areas with low or oscillating WSS which are more evident at the lesion models with stenosis severity of 70$\%$ were observed as shown in Fig.\ref{fig:wss_captions}.

The oscillatory patterns together with the secondary flow that occurs inside the stenosis and especially at regions with high curvature, have a significant contribution to the RBC deposition in the lesion. These changes downstream the maximum curvature position can potentially lead to a reduced WSS. In the areas of lower WSS, the endothelium might be more easily compromised by other particulates (monocytes and proteins) which might cause thrombosis and atherogenesis \cite{melder1995selectin, rajendran2013vascular}. 

\section{Conclusion} 
The paper here presents a novel study for the impact of hematocrit for calculating pressure drop through eccentric tubular lesions considering the most important fluid forces such as lift, drag, virtual, gravity, which has not been previously presented in the literature. 
{\color{black} Our results showed that} pressure drop is not strongly affected from single phase, and neither is mean velocity {\color{black} thus confirming single phase modelling is adequate for estimating these variables. This implies the validity of the majority of previous studies.} On the other hand, particle buildup significantly changes which is only revealed by the multiphase approach. 
The observed migration of red blood cells and their interactions at the surface of endothelial cells are usually ignored in single-phase algorithms. With the present approach, considering additional blood components, one may get a more accurate view of the wall shear stress and particulate buildup related to cardiovascular diseases such as atherogenesis.
Hydrodynamic diffusion of RBCs occurs downstream the stenosis and is more intensive for higher stenosis severity changing the RBC distribution and WSS.

Future studies will include the effect of these extra particulates in the model since the model has no limitations in the number of phases considered. Moreover, although a lot of efford has been made to consider a plethora of different sub-models for the occuring physical phenomena, the present work could be further improved without being limited to the assumption of spherical RBC. Options for allowing nonspherical particles using the shape factor concept \cite{jung2006hemodynamic} could provide more insights for the haemodynamic parameters studied here. 

\section*{Conflict of Interest} 
The authors declare no conflict of interest.
 
\section*{Acknowledgment}
This work was supported by Heart Research UK [RG2670/18/21] and the Wellcome/EPSRC Centre for Medical Engineering [WT 203148/Z/16/Z]. 

\section*{References}

\bibliographystyle{model6-num-names}
\bibliography{mybibfile}

\begin{thebibliography}{56}
\expandafter\ifx\csname natexlab\endcsname\relax\def\natexlab#1{#1}\fi
\providecommand{\url}[1]{\texttt{#1}}
\providecommand{\href}[2]{#2}
\providecommand{\path}[1]{#1}
\providecommand{\DOIprefix}{doi:}
\providecommand{\ArXivprefix}{arXiv:}
\providecommand{\URLprefix}{URL: }
\providecommand{\Pubmedprefix}{pmid:}
\providecommand{\doi}[1]{\href{http://dx.doi.org/#1}{\path{#1}}}
\providecommand{\Pubmed}[1]{\href{pmid:#1}{\path{#1}}}
\providecommand{\bibinfo}[2]{#2}
\ifx\xfnm\relax \def\xfnm[#1]{\unskip,\space#1}\fi
\bibitem[{Reinhart(2016)}]{reinhart2016optimum}
\bibinfo{author}{W.~H. Reinhart},
\newblock \bibinfo{title}{The optimum hematocrit},
\newblock \bibinfo{journal}{Clinical hemorheology and microcirculation}
  \bibinfo{volume}{64} (\bibinfo{year}{2016}) \bibinfo{pages}{575--585}.
\bibitem[{Nader et~al.(2019)Nader, Skinner, Romana, Fort, Lemonne, Guillot,
  Gauthier, Antoine-Jonville, Renoux, Hardy-Dessources et~al.}]{nader2019blood}
\bibinfo{author}{E.~Nader}, \bibinfo{author}{S.~Skinner},
  \bibinfo{author}{M.~Romana}, \bibinfo{author}{R.~Fort},
  \bibinfo{author}{N.~Lemonne}, \bibinfo{author}{N.~Guillot},
  \bibinfo{author}{A.~Gauthier}, \bibinfo{author}{S.~Antoine-Jonville},
  \bibinfo{author}{C.~Renoux}, \bibinfo{author}{M.-D. Hardy-Dessources},
  et~al.,
\newblock \bibinfo{title}{Blood rheology: key parameters, impact on blood flow,
  role in sickle cell disease and effects of exercise},
\newblock \bibinfo{journal}{Frontiers in physiology} \bibinfo{volume}{10}
  (\bibinfo{year}{2019}) \bibinfo{pages}{1329}.
\bibitem[{Mehri et~al.(2018)Mehri, Mavriplis, and Fenech}]{mehri2018red}
\bibinfo{author}{R.~Mehri}, \bibinfo{author}{C.~Mavriplis},
  \bibinfo{author}{M.~Fenech},
\newblock \bibinfo{title}{Red blood cell aggregates and their effect on
  non-newtonian blood viscosity at low hematocrit in a two-fluid low shear rate
  microfluidic system},
\newblock \bibinfo{journal}{Plos one} \bibinfo{volume}{13}
  (\bibinfo{year}{2018}) \bibinfo{pages}{e0199911}.
\bibitem[{Yin et~al.(2013)Yin, Thomas, and Zhang}]{yin2013multiple}
\bibinfo{author}{X.~Yin}, \bibinfo{author}{T.~Thomas},
  \bibinfo{author}{J.~Zhang},
\newblock \bibinfo{title}{Multiple red blood cell flows through microvascular
  bifurcations: cell free layer, cell trajectory, and hematocrit separation},
\newblock \bibinfo{journal}{Microvascular research} \bibinfo{volume}{89}
  (\bibinfo{year}{2013}) \bibinfo{pages}{47--56}.
\bibitem[{Kim et~al.(2016)Kim, Antaki, and Massoudi}]{kim2016computational}
\bibinfo{author}{J.~Kim}, \bibinfo{author}{J.~F. Antaki},
  \bibinfo{author}{M.~Massoudi},
\newblock \bibinfo{title}{Computational study of blood flow in microchannels},
\newblock \bibinfo{journal}{Journal of computational and applied mathematics}
  \bibinfo{volume}{292} (\bibinfo{year}{2016}) \bibinfo{pages}{174--187}.
\bibitem[{Hirt and Nichols(1981)}]{Hirt1981}
\bibinfo{author}{C.~W. Hirt}, \bibinfo{author}{B.~D. Nichols},
\newblock \bibinfo{title}{Volume of fluid (vof) method for the dynamics of free
  boundaries},
\newblock \bibinfo{journal}{Journal of Computational Physics}
  \bibinfo{volume}{39} (\bibinfo{year}{1981}) \bibinfo{pages}{201 -- 225}.
\bibitem[{Prosperetti and Tryggvason(2009)}]{Prosperetti2009}
\bibinfo{author}{A.~Prosperetti}, \bibinfo{author}{G.~Tryggvason},
  \bibinfo{title}{Computational Methods for Multiphase Flow},
  \bibinfo{publisher}{Cambridge University Press}, \bibinfo{year}{2009}.
\bibitem[{Lyras and Lee(2021)}]{LyrasLee2021a}
\bibinfo{author}{K.~Lyras}, \bibinfo{author}{J.~Lee},
\newblock \bibinfo{title}{A finite volume coupled level set and volume of fluid
  method with a mass conservation step for simulating two-phase flows},
\newblock \bibinfo{journal}{International Journal for Numerical Methods in
  Fluids} \bibinfo{volume}{X} (\bibinfo{year}{2021}) \bibinfo{pages}{In press}.
\bibitem[{Hausmann et~al.(1994)Hausmann, Lundkvist, Friedrich, Sudhir,
  Fitzgerald, and Yock}]{hausmann1994lumen}
\bibinfo{author}{D.~Hausmann}, \bibinfo{author}{A.~J. Lundkvist},
  \bibinfo{author}{G.~Friedrich}, \bibinfo{author}{K.~Sudhir},
  \bibinfo{author}{P.~J. Fitzgerald}, \bibinfo{author}{P.~G. Yock},
\newblock \bibinfo{title}{Lumen and plaque shape in atherosclerotic coronary
  arteries assessed by in vivo intracoronary ultrasound},
\newblock \bibinfo{journal}{The American journal of cardiology}
  \bibinfo{volume}{74} (\bibinfo{year}{1994}) \bibinfo{pages}{857--863}.
\bibitem[{Waller(1989)}]{waller1989eccentric}
\bibinfo{author}{B.~Waller},
\newblock \bibinfo{title}{The eccentric coronary atherosclerotic plaque:
  morphologic observations and clinical relevance},
\newblock \bibinfo{journal}{Clinical cardiology} \bibinfo{volume}{12}
  (\bibinfo{year}{1989}) \bibinfo{pages}{14--20}.
\bibitem[{Kaazempur-Mofrad et~al.(2005)Kaazempur-Mofrad, Wada, Myers, and
  Ethier}]{kaazempur2005mass}
\bibinfo{author}{M.~Kaazempur-Mofrad}, \bibinfo{author}{S.~Wada},
  \bibinfo{author}{J.~Myers}, \bibinfo{author}{C.~Ethier},
\newblock \bibinfo{title}{Mass transport and fluid flow in stenotic arteries:
  axisymmetric and asymmetric models},
\newblock \bibinfo{journal}{International Journal of Heat and Mass Transfer}
  \bibinfo{volume}{48} (\bibinfo{year}{2005}) \bibinfo{pages}{4510--4517}.
\bibitem[{Lyras and Lee(2021)}]{LyrasLee2021c}
\bibinfo{author}{K.~Lyras}, \bibinfo{author}{J.~Lee},
\newblock \bibinfo{title}{Comparison of numerical implementations for modelling
  flow through arterial stenoses},
\newblock \bibinfo{journal}{International Journal of Mechanical Sciences}
  \bibinfo{volume}{X} (\bibinfo{year}{2021}) \bibinfo{pages}{In press}.
\bibitem[{Gholipour et~al.(2018)Gholipour, Ghayesh, Zander, and
  Mahajan}]{gholipour2018}
\bibinfo{author}{A.~Gholipour}, \bibinfo{author}{M.~H. Ghayesh},
  \bibinfo{author}{A.~Zander}, \bibinfo{author}{R.~Mahajan},
\newblock \bibinfo{title}{Three-dimensional biomechanics of coronary arteries},
\newblock \bibinfo{journal}{International Journal of Engineering Science}
  \bibinfo{volume}{130} (\bibinfo{year}{2018}) \bibinfo{pages}{93--114}.
\bibitem[{Zhao et~al.(2021)Zhao, Vatankhah, Goh, Michelis, Kyanian, Zhang, Li,
  and Ju}]{zhao2021}
\bibinfo{author}{Y.~C. Zhao}, \bibinfo{author}{P.~Vatankhah},
  \bibinfo{author}{T.~Goh}, \bibinfo{author}{R.~Michelis},
  \bibinfo{author}{K.~Kyanian}, \bibinfo{author}{Y.~Zhang},
  \bibinfo{author}{Z.~Li}, \bibinfo{author}{L.~A. Ju},
\newblock \bibinfo{title}{Hemodynamic analysis for stenosis microfluidic model
  of thrombosis with refined computational fluid dynamics simulation},
\newblock \bibinfo{journal}{Scientific reports} \bibinfo{volume}{11}
  (\bibinfo{year}{2021}) \bibinfo{pages}{1--10}.
\bibitem[{Kushner et~al.(2009)Kushner, Hand, Smith, King, Anderson, Antman,
  Bailey, Bates, Blankenship, Casey et~al.}]{kushner2009}
\bibinfo{author}{F.~G. Kushner}, \bibinfo{author}{M.~Hand},
  \bibinfo{author}{S.~C. Smith}, \bibinfo{author}{S.~B. King},
  \bibinfo{author}{J.~L. Anderson}, \bibinfo{author}{E.~M. Antman},
  \bibinfo{author}{S.~R. Bailey}, \bibinfo{author}{E.~R. Bates},
  \bibinfo{author}{J.~C. Blankenship}, \bibinfo{author}{D.~E. Casey}, et~al.,
\newblock \bibinfo{title}{2009 focused updates: Acc/aha guidelines for the
  management of patients with st-elevation myocardial infarction (updating the
  2004 guideline and 2007 focused update) and acc/aha/scai guidelines on
  percutaneous coronary intervention (updating the 2005 guideline and 2007
  focused update) a report of the american college of cardiology
  foundation/american heart association task force on practice guidelines},
\newblock \bibinfo{journal}{Journal of the American College of Cardiology}
  \bibinfo{volume}{54} (\bibinfo{year}{2009}) \bibinfo{pages}{2205--2241}.
\bibitem[{Benis et~al.(1970)Benis, Usami, and CHIEN}]{benis1970effect}
\bibinfo{author}{A.~M. Benis}, \bibinfo{author}{S.~Usami},
  \bibinfo{author}{S.~CHIEN},
\newblock \bibinfo{title}{Effect of hematocrit and inertial losses on
  pressure-flow relations in the isolated hindpaw of the dog},
\newblock \bibinfo{journal}{Circulation research} \bibinfo{volume}{27}
  (\bibinfo{year}{1970}) \bibinfo{pages}{1047--1068}.
\bibitem[{{\c{C}}{\i}nar et~al.(1999){\c{C}}{\i}nar, Demir, Pa{\c{c}}, and
  {\c{C}}{\i}nar}]{ccinar1999effect}
\bibinfo{author}{Y.~{\c{C}}{\i}nar}, \bibinfo{author}{G.~Demir},
  \bibinfo{author}{M.~Pa{\c{c}}}, \bibinfo{author}{A.~B. {\c{C}}{\i}nar},
\newblock \bibinfo{title}{Effect of hematocrit on blood pressure via
  hyperviscosity},
\newblock \bibinfo{journal}{American journal of hypertension}
  \bibinfo{volume}{12} (\bibinfo{year}{1999}) \bibinfo{pages}{739--743}.
\bibitem[{Jung et~al.(2006)Jung, Hassanein, and
  Lyczkowski}]{jung2006hemodynamic}
\bibinfo{author}{J.~Jung}, \bibinfo{author}{A.~Hassanein},
  \bibinfo{author}{R.~W. Lyczkowski},
\newblock \bibinfo{title}{Hemodynamic computation using multiphase flow
  dynamics in a right coronary artery},
\newblock \bibinfo{journal}{Annals of biomedical engineering}
  \bibinfo{volume}{34} (\bibinfo{year}{2006}) \bibinfo{pages}{393--407}.
\bibitem[{Kim et~al.(2008)Kim, VandeVord, and Lee}]{kim2008multiphase}
\bibinfo{author}{Y.~H. Kim}, \bibinfo{author}{P.~J. VandeVord},
  \bibinfo{author}{J.~S. Lee},
\newblock \bibinfo{title}{Multiphase non-newtonian effects on pulsatile
  hemodynamics in a coronary artery},
\newblock \bibinfo{journal}{International journal for numerical methods in
  fluids} \bibinfo{volume}{58} (\bibinfo{year}{2008})
  \bibinfo{pages}{803--825}.
\bibitem[{Baskurt et~al.(2011)Baskurt, Neu, and Meiselman}]{baskurt2011red}
\bibinfo{author}{O.~Baskurt}, \bibinfo{author}{B.~Neu}, \bibinfo{author}{H.~J.
  Meiselman},
\newblock \bibinfo{title}{Red blood cell aggregation}  (\bibinfo{year}{2011}).
\bibitem[{L{\"u}cker et~al.(2015)L{\"u}cker, Weber, and
  Jenny}]{lucker2015dynamic}
\bibinfo{author}{A.~L{\"u}cker}, \bibinfo{author}{B.~Weber},
  \bibinfo{author}{P.~Jenny},
\newblock \bibinfo{title}{A dynamic model of oxygen transport from capillaries
  to tissue with moving red blood cells},
\newblock \bibinfo{journal}{American Journal of Physiology-Heart and
  Circulatory Physiology} \bibinfo{volume}{308} (\bibinfo{year}{2015})
  \bibinfo{pages}{H206--H216}.
\bibitem[{Ling et~al.(2021)Ling, Tang, and Liu}]{ling2021numerical}
\bibinfo{author}{Y.~Ling}, \bibinfo{author}{J.~Tang}, \bibinfo{author}{H.~Liu},
\newblock \bibinfo{title}{Numerical investigation of two-phase non-newtonian
  blood flow in bifurcate pulmonary arteries with a flow resistant using
  eulerian multiphase model},
\newblock \bibinfo{journal}{Chemical Engineering Science} \bibinfo{volume}{233}
  (\bibinfo{year}{2021}) \bibinfo{pages}{116426}.
\bibitem[{Thim et~al.(2012)Thim, Hagensen, Falk, H{\o}rlyck, Kim, Niemann,
  Thrys{\o}e, Drouet, Paaske, and B{\o}tker}]{Thim2012}
\bibinfo{author}{T.~Thim}, \bibinfo{author}{M.~K. Hagensen},
  \bibinfo{author}{E.~Falk}, \bibinfo{author}{A.~H{\o}rlyck},
  \bibinfo{author}{W.~Y. Kim}, \bibinfo{author}{A.~K. Niemann},
  \bibinfo{author}{S.~A. Thrys{\o}e}, \bibinfo{author}{L.~Drouet},
  \bibinfo{author}{W.~P. Paaske}, \bibinfo{author}{H.~E. B{\o}tker},
\newblock \bibinfo{title}{Wall shear stress and local plaque development in
  stenosed carotid arteries of hypercholesterolemic minipigs},
\newblock \bibinfo{journal}{Journal of cardiovascular disease research}
  \bibinfo{volume}{3} (\bibinfo{year}{2012}) \bibinfo{pages}{76--83}.
\bibitem[{Han et~al.(2016)Han, Starikov, {\'o}~Hartaigh, Gransar, Kolli, Lee,
  Rizvi, Baskaran, Schulman-Marcus, Lin et~al.}]{Han2016}
\bibinfo{author}{D.~Han}, \bibinfo{author}{A.~Starikov},
  \bibinfo{author}{B.~{\'o}~Hartaigh}, \bibinfo{author}{H.~Gransar},
  \bibinfo{author}{K.~K. Kolli}, \bibinfo{author}{J.~H. Lee},
  \bibinfo{author}{A.~Rizvi}, \bibinfo{author}{L.~Baskaran},
  \bibinfo{author}{J.~Schulman-Marcus}, \bibinfo{author}{F.~Y. Lin}, et~al.,
\newblock \bibinfo{title}{Relationship between endothelial wall shear stress
  and high-risk atherosclerotic plaque characteristics for identification of
  coronary lesions that cause ischemia: A direct comparison with fractional
  flow reserve},
\newblock \bibinfo{journal}{Journal of the American Heart Association}
  \bibinfo{volume}{5} (\bibinfo{year}{2016}) \bibinfo{pages}{e004186}.
\bibitem[{Weller et~al.(1998)Weller, Tabor, Jasak, and Fureby}]{weller1998}
\bibinfo{author}{H.~G. Weller}, \bibinfo{author}{G.~Tabor},
  \bibinfo{author}{H.~Jasak}, \bibinfo{author}{C.~Fureby},
\newblock \bibinfo{title}{A tensorial approach to computational continuum
  mechanics using object-oriented techniques},
\newblock \bibinfo{journal}{Comput. Phys.} \bibinfo{volume}{12}
  (\bibinfo{year}{1998}) \bibinfo{pages}{620--631}.
\bibitem[{Wardle and Weller(2013)}]{wardle2013hybrid}
\bibinfo{author}{K.~E. Wardle}, \bibinfo{author}{H.~G. Weller},
\newblock \bibinfo{title}{Hybrid multiphase cfd solver for coupled
  dispersed/segregated flows in liquid-liquid extraction},
\newblock \bibinfo{journal}{International Journal of Chemical Engineering}
  \bibinfo{volume}{2013} (\bibinfo{year}{2013}).
\bibitem[{Schiller(1933)}]{schiller1933drag}
\bibinfo{author}{L.~Schiller},
\newblock \bibinfo{title}{A drag coefficient correlation},
\newblock \bibinfo{journal}{Zeit. Ver. Deutsch. Ing.} \bibinfo{volume}{77}
  (\bibinfo{year}{1933}) \bibinfo{pages}{318--320}.
\bibitem[{Olsson and Kreiss(2005)}]{Olsson2005}
\bibinfo{author}{E.~Olsson}, \bibinfo{author}{G.~Kreiss},
\newblock \bibinfo{title}{A conservative level set method for two phase flow},
\newblock \bibinfo{journal}{Journal of Computational Physics}
  \bibinfo{volume}{210} (\bibinfo{year}{2005}) \bibinfo{pages}{225--246}.
\bibitem[{Kees et~al.(2011)Kees, Akkerman, Farthing, and Bazilevs}]{Kees2011}
\bibinfo{author}{C.~E. Kees}, \bibinfo{author}{I.~Akkerman},
  \bibinfo{author}{M.~W. Farthing}, \bibinfo{author}{Y.~Bazilevs},
\newblock \bibinfo{title}{A conservative level set method suitable for
  variable-order approximations and unstructured meshes},
\newblock \bibinfo{journal}{J. Comput. Phys.} \bibinfo{volume}{230}
  (\bibinfo{year}{2011}) \bibinfo{pages}{4536--4558}.
\bibitem[{Kunkelmann and Stephan(2010)}]{Kunkelmann2010}
\bibinfo{author}{C.~Kunkelmann}, \bibinfo{author}{P.~Stephan},
  \bibinfo{title}{Modification and extension of a standard volume-of-fluid
  solverfor simulating boiling heat transfer}, \bibinfo{publisher}{ECCOMAS
  CFD2010}, \bibinfo{year}{2010}.
\bibitem[{Albadawi et~al.(2013)Albadawi, Donoghue, Robinson, Murray, and
  Delauré}]{Albadawi2013}
\bibinfo{author}{A.~Albadawi}, \bibinfo{author}{D.~B. Donoghue},
  \bibinfo{author}{A.~J. Robinson}, \bibinfo{author}{D.~B. Murray},
  \bibinfo{author}{Y.~M.~C. Delauré},
\newblock \bibinfo{title}{Influence of surface tension implementation in volume
  of fluid and coupled volume of fluid with level set methods for bubble growth
  and detachment},
\newblock \bibinfo{journal}{International Journal of Multiphase Flow}
  \bibinfo{volume}{53} (\bibinfo{year}{2013}) \bibinfo{pages}{11 -- 28}.
\bibitem[{Dianat et~al.(2017)Dianat, Skarysz, and Garmory}]{Dianat2017}
\bibinfo{author}{M.~Dianat}, \bibinfo{author}{M.~Skarysz},
  \bibinfo{author}{A.~Garmory},
\newblock \bibinfo{title}{A coupled level set and volume of fluid method for
  automotive exterior water management applications},
\newblock \bibinfo{journal}{International Journal of Multiphase Flow}
  \bibinfo{volume}{91} (\bibinfo{year}{2017}) \bibinfo{pages}{19 -- 38}.
\bibitem[{Lyras et~al.(2020)Lyras, Hanson, Fairweather, and Heggs}]{Lyras2020}
\bibinfo{author}{K.~G. Lyras}, \bibinfo{author}{B.~Hanson},
  \bibinfo{author}{M.~Fairweather}, \bibinfo{author}{P.~J. Heggs},
\newblock \bibinfo{title}{A coupled level set and volume of fluid method with a
  re-initialisation step suitable for unstructured meshes},
\newblock \bibinfo{journal}{Journal of Computational Physics}
  \bibinfo{volume}{407} (\bibinfo{year}{2020}) \bibinfo{pages}{109224}.
\bibitem[{Hartmann et~al.(2010)Hartmann, Meinke, and
  Schr{\"o}der}]{Hartmann2010}
\bibinfo{author}{D.~Hartmann}, \bibinfo{author}{M.~Meinke},
  \bibinfo{author}{W.~Schr{\"o}der},
\newblock \bibinfo{title}{The constrained reinitialization equation for level
  set methods},
\newblock \bibinfo{journal}{Journal of Computational Physics}
  \bibinfo{volume}{229} (\bibinfo{year}{2010}) \bibinfo{pages}{1514--1535}.
\bibitem[{Young and Tsai(1973{\natexlab{a}})}]{young1973steady}
\bibinfo{author}{D.~F. Young}, \bibinfo{author}{F.~Y. Tsai},
\newblock \bibinfo{title}{Flow characteristics in models of arterial
  stenoses—i. steady flow},
\newblock \bibinfo{journal}{Journal of biomechanics} \bibinfo{volume}{6}
  (\bibinfo{year}{1973}{\natexlab{a}}) \bibinfo{pages}{395--410}.
\bibitem[{Young and Tsai(1973{\natexlab{b}})}]{young1973unsteady}
\bibinfo{author}{D.~F. Young}, \bibinfo{author}{F.~Y. Tsai},
\newblock \bibinfo{title}{Flow characteristics in models of arterial
  stenoses—ii. unsteady flow},
\newblock \bibinfo{journal}{Journal of biomechanics} \bibinfo{volume}{6}
  (\bibinfo{year}{1973}{\natexlab{b}}) \bibinfo{pages}{547--559}.
\bibitem[{Lyras and Lee(2021)}]{LyrasLee2021b}
\bibinfo{author}{K.~Lyras}, \bibinfo{author}{J.~Lee},
\newblock \bibinfo{title}{Numerical investigation of reduced-order modelling in
  arterial stenosis},
\newblock \bibinfo{journal}{PLOS ONE} \bibinfo{volume}{X}
  (\bibinfo{year}{2021}) \bibinfo{pages}{In press}.
\bibitem[{Stergiopulos et~al.(1992)Stergiopulos, Young, and
  Rogge}]{stergiopulos1992}
\bibinfo{author}{N.~Stergiopulos}, \bibinfo{author}{D.~Young},
  \bibinfo{author}{T.~Rogge},
\newblock \bibinfo{title}{Computer simulation of arterial flow with
  applications to arterial and aortic stenoses},
\newblock \bibinfo{journal}{Journal of biomechanics} \bibinfo{volume}{25}
  (\bibinfo{year}{1992}) \bibinfo{pages}{1477--1488}.
\bibitem[{Menter(1993)}]{menter1993}
\bibinfo{author}{F.~Menter},
\newblock \bibinfo{title}{Zonal two equation k-w turbulence models for
  aerodynamic flows},
\newblock in: \bibinfo{booktitle}{23rd fluid dynamics, plasmadynamics, and
  lasers conference}, \bibinfo{year}{1993}, p. \bibinfo{pages}{2906}.
\bibitem[{Khair et~al.(2015)Khair, Wang, and Kuhn}]{khair2015}
\bibinfo{author}{A.~Khair}, \bibinfo{author}{B.-C. Wang},
  \bibinfo{author}{D.~C. Kuhn},
\newblock \bibinfo{title}{Study of laminar--turbulent flow transition under
  pulsatile conditions in a constricted channel},
\newblock \bibinfo{journal}{International Journal of Computational Fluid
  Dynamics} \bibinfo{volume}{29} (\bibinfo{year}{2015})
  \bibinfo{pages}{447--463}.
\bibitem[{Wu et~al.(2015)Wu, Yang, Antaki, Aubry, and Massoudi}]{wu2015study}
\bibinfo{author}{W.-T. Wu}, \bibinfo{author}{F.~Yang}, \bibinfo{author}{J.~F.
  Antaki}, \bibinfo{author}{N.~Aubry}, \bibinfo{author}{M.~Massoudi},
\newblock \bibinfo{title}{Study of blood flow in several benchmark
  micro-channels using a two-fluid approach},
\newblock \bibinfo{journal}{International journal of engineering science}
  \bibinfo{volume}{95} (\bibinfo{year}{2015}) \bibinfo{pages}{49--59}.
\bibitem[{Xu et~al.(2013)Xu, Kaliviotis, Munjiza, Avital, Ji, and
  Williams}]{xu2013large}
\bibinfo{author}{D.~Xu}, \bibinfo{author}{E.~Kaliviotis},
  \bibinfo{author}{A.~Munjiza}, \bibinfo{author}{E.~Avital},
  \bibinfo{author}{C.~Ji}, \bibinfo{author}{J.~Williams},
\newblock \bibinfo{title}{Large scale simulation of red blood cell aggregation
  in shear flows},
\newblock \bibinfo{journal}{Journal of biomechanics} \bibinfo{volume}{46}
  (\bibinfo{year}{2013}) \bibinfo{pages}{1810--1817}.
\bibitem[{Freidoonimehr et~al.(2018)Freidoonimehr, Arjomandi, Chin, and
  Zander}]{freidoonimehr2018effect}
\bibinfo{author}{N.~Freidoonimehr}, \bibinfo{author}{M.~Arjomandi},
  \bibinfo{author}{R.~Chin}, \bibinfo{author}{A.~Zander},
\newblock \bibinfo{title}{Effect of degree of stenosis on the pulsatile flow
  pressure drop in a coronary artery}  (\bibinfo{year}{2018}).
\bibitem[{Gould(1978)}]{gould1978}
\bibinfo{author}{K.~L. Gould},
\newblock \bibinfo{title}{Pressure-flow characteristics of coronary stenoses in
  unsedated dogs at rest and during coronary vasodilation.},
\newblock \bibinfo{journal}{Circulation research} \bibinfo{volume}{43}
  (\bibinfo{year}{1978}) \bibinfo{pages}{242--253}.
\bibitem[{Buradi et~al.(2019)Buradi, Morab, and Mahalingam}]{buradi2019effect}
\bibinfo{author}{A.~Buradi}, \bibinfo{author}{S.~Morab},
  \bibinfo{author}{A.~Mahalingam},
\newblock \bibinfo{title}{Effect of stenosis severity on shear-induced
  diffusion of red blood cells in coronary arteries},
\newblock \bibinfo{journal}{Journal of Mechanics in Medicine and Biology}
  \bibinfo{volume}{19} (\bibinfo{year}{2019}) \bibinfo{pages}{1950034}.
\bibitem[{Brooks et~al.(1970)Brooks, Goodwin, and
  Seaman}]{brooks1970interactions}
\bibinfo{author}{D.~Brooks}, \bibinfo{author}{J.~Goodwin},
  \bibinfo{author}{G.~Seaman},
\newblock \bibinfo{title}{Interactions among erythrocytes under shear.},
\newblock \bibinfo{journal}{Journal of Applied Physiology} \bibinfo{volume}{28}
  (\bibinfo{year}{1970}) \bibinfo{pages}{172--177}.
\bibitem[{Massoudi et~al.(2012)Massoudi, Kim, and
  Antaki}]{massoudi2012modeling}
\bibinfo{author}{M.~Massoudi}, \bibinfo{author}{J.~Kim}, \bibinfo{author}{J.~F.
  Antaki},
\newblock \bibinfo{title}{Modeling and numerical simulation of blood flow using
  the theory of interacting continua},
\newblock \bibinfo{journal}{International journal of non-linear mechanics}
  \bibinfo{volume}{47} (\bibinfo{year}{2012}) \bibinfo{pages}{506--520}.
\bibitem[{Yeleswarapu(1996)}]{yeleswarapu1996evaluation}
\bibinfo{author}{K.~K. Yeleswarapu}, \bibinfo{title}{Evaluation of continuum
  models for characterizing the constitutive behavior of blood},
  \bibinfo{publisher}{University of Pittsburgh}, \bibinfo{year}{1996}.
\bibitem[{Zhao and Shaqfeh(2011)}]{zhao2011shear}
\bibinfo{author}{H.~Zhao}, \bibinfo{author}{E.~S. Shaqfeh},
\newblock \bibinfo{title}{Shear-induced platelet margination in a
  microchannel},
\newblock \bibinfo{journal}{Physical Review E} \bibinfo{volume}{83}
  (\bibinfo{year}{2011}) \bibinfo{pages}{061924}.
\bibitem[{Ku et~al.(1985)Ku, Giddens, Zarins, and Glagov}]{ku1985pulsatile}
\bibinfo{author}{D.~N. Ku}, \bibinfo{author}{D.~P. Giddens},
  \bibinfo{author}{C.~K. Zarins}, \bibinfo{author}{S.~Glagov},
\newblock \bibinfo{title}{Pulsatile flow and atherosclerosis in the human
  carotid bifurcation. positive correlation between plaque location and low
  oscillating shear stress.},
\newblock \bibinfo{journal}{Arteriosclerosis: An Official Journal of the
  American Heart Association, Inc.} \bibinfo{volume}{5} (\bibinfo{year}{1985})
  \bibinfo{pages}{293--302}.
\bibitem[{Mahalingam et~al.(2016)Mahalingam, Gawandalkar, Kini, Buradi, Araki,
  Ikeda, Nicolaides, Laird, Saba, and Suri}]{mahalingam2016numerical}
\bibinfo{author}{A.~Mahalingam}, \bibinfo{author}{U.~U. Gawandalkar},
  \bibinfo{author}{G.~Kini}, \bibinfo{author}{A.~Buradi},
  \bibinfo{author}{T.~Araki}, \bibinfo{author}{N.~Ikeda},
  \bibinfo{author}{A.~Nicolaides}, \bibinfo{author}{J.~R. Laird},
  \bibinfo{author}{L.~Saba}, \bibinfo{author}{J.~S. Suri},
\newblock \bibinfo{title}{Numerical analysis of the effect of turbulence
  transition on the hemodynamic parameters in human coronary arteries},
\newblock \bibinfo{journal}{Cardiovascular Diagnosis And Therapy}
  \bibinfo{volume}{6} (\bibinfo{year}{2016}) \bibinfo{pages}{208}.
\bibitem[{Ogawa et~al.(1990)Ogawa, Shreeniwas, Brett, Clauss, Furie, and
  Stern}]{ogawa1990effect}
\bibinfo{author}{S.~Ogawa}, \bibinfo{author}{R.~Shreeniwas},
  \bibinfo{author}{J.~Brett}, \bibinfo{author}{M.~Clauss},
  \bibinfo{author}{M.~Furie}, \bibinfo{author}{D.~M. Stern},
\newblock \bibinfo{title}{The effect of hypoxia on capillary endothelial cell
  function: modulation of barrier and coagulant function},
\newblock \bibinfo{journal}{British journal of haematology}
  \bibinfo{volume}{75} (\bibinfo{year}{1990}) \bibinfo{pages}{517--524}.
\bibitem[{Premont et~al.(2020)Premont, Reynolds, Zhang, and
  Stamler}]{premont2020role}
\bibinfo{author}{R.~T. Premont}, \bibinfo{author}{J.~D. Reynolds},
  \bibinfo{author}{R.~Zhang}, \bibinfo{author}{J.~S. Stamler},
\newblock \bibinfo{title}{Role of nitric oxide carried by hemoglobin in
  cardiovascular physiology: developments on a three-gas respiratory cycle},
\newblock \bibinfo{journal}{Circulation research} \bibinfo{volume}{126}
  (\bibinfo{year}{2020}) \bibinfo{pages}{129--158}.
\bibitem[{Jeney et~al.(2014)Jeney, Balla, and Balla}]{jeney2014red}
\bibinfo{author}{V.~Jeney}, \bibinfo{author}{G.~Balla},
  \bibinfo{author}{J.~Balla},
\newblock \bibinfo{title}{Red blood cell, hemoglobin and heme in the
  progression of atherosclerosis},
\newblock \bibinfo{journal}{Frontiers in physiology} \bibinfo{volume}{5}
  (\bibinfo{year}{2014}) \bibinfo{pages}{379}.
\bibitem[{Melder et~al.(1995)Melder, Munn, Yamada, Ohkubo, and
  Jain}]{melder1995selectin}
\bibinfo{author}{R.~Melder}, \bibinfo{author}{L.~Munn},
  \bibinfo{author}{S.~Yamada}, \bibinfo{author}{C.~Ohkubo},
  \bibinfo{author}{R.~Jain},
\newblock \bibinfo{title}{Selectin-and integrin-mediated t-lymphocyte rolling
  and arrest on tnf-alpha-activated endothelium: augmentation by erythrocytes},
\newblock \bibinfo{journal}{Biophysical journal} \bibinfo{volume}{69}
  (\bibinfo{year}{1995}) \bibinfo{pages}{2131--2138}.
\bibitem[{Rajendran et~al.(2013)Rajendran, Rengarajan, Thangavel, Nishigaki,
  Sakthisekaran, Sethi, and Nishigaki}]{rajendran2013vascular}
\bibinfo{author}{P.~Rajendran}, \bibinfo{author}{T.~Rengarajan},
  \bibinfo{author}{J.~Thangavel}, \bibinfo{author}{Y.~Nishigaki},
  \bibinfo{author}{D.~Sakthisekaran}, \bibinfo{author}{G.~Sethi},
  \bibinfo{author}{I.~Nishigaki},
\newblock \bibinfo{title}{The vascular endothelium and human diseases},
\newblock \bibinfo{journal}{International journal of biological sciences}
  \bibinfo{volume}{9} (\bibinfo{year}{2013}) \bibinfo{pages}{1057}.

\end{thebibliography}

\end{document}